\RequirePackage[2020-02-02]{latexrelease}
\documentclass{iau}

\usepackage{amsmath}
\usepackage{graphicx}
\usepackage{multirow}
\usepackage{MnSymbol}
\usepackage{xspace}
\usepackage{natbib}
\usepackage[utf8]{inputenc}

\begin{document}

\def\eddr{$\alpha_{\rm Edd}$}
\def\mbh{$M_{\bullet}$}
\def\eddr{$\alpha_{\rm Edd}$}
\def\civ{{C\sc{iv}} $\lambda$1549\/}
\def\mgii{{Mg\sc{ii}} $\lambda$2800\/}
\def\civa{{C\sc{iv}}~}
\def\mgiia{{Mg\sc{ii}}~}
\def\feii{{Fe\sc{ii}}~}
\def\Mo{{M$_{\odot}$}}

\lefttitle{2.5D FRADO model in BAL QSOs}
\righttitle{Proceedings of the International Astronomical Union: \LaTeX\ Guidelines for~authors}

\jnlPage{1}{7}
\jnlDoiYr{2023}
\doival{10.1017/xxxxx}

\aopheadtitle{Proceedings IAU Symposium}
\editors{C. Sterken,  J. Hearnshaw \&  D. Valls-Gabaud, eds.}

\title{BAL effect in quasars due to source orientation}

\author{M. Sniegowska$^{1,2}$, M. H. Naddaf$^{1,2}$\footnote{naddaf@cft.edu.pl}, M. L. Martinez-Aldama$^{3,4}$,\\ P. Marziani$^{5}$, S. Panda$^{6}$\footnote{CNPq Fellow}, B. Czerny$^{1}$}

\affiliation{$^{1}$Center for Theoretical Physics, Polish Academy of  Sciences, Lotnikow 32/46, 02-668 Warsaw, Poland}

\affiliation{$^{2}$Nicolaus Copernicus Astronomical Center, Polish Academy of Sciences, Bartycka 18, 00-716 Warsaw, Poland}

\affiliation{$^{3}$Instituto de Física y Astronomía, Facultad de Ciencias, Universidad de Valparaíso, Gran Bretaña 1111, Valparaíso, Chile}

\affiliation{$^{4}$Departamento de Astronomía, Universidad de Chile, Casilla 36D, Santiago, Chile}

\affiliation{$^{5}$INAF-Astronomical Observatory of Padova, Vicolo dell'Osservatorio, 5, 35122 Padova PD, Italy}

\affiliation{$^{6}$Laborat\'orio Nacional de Astrof\'isica - Rua dos Estados Unidos 154, Bairro das Na\c c\~oes. CEP 37504-364, Itajub\'a, MG, Brazil}

\begin{abstract}
We investigated a scenario where the presence of a broad absorption line (BAL) feature in quasars (QSOs) is contingent upon the line of sight being situated within an outflow cone emanating from the source. We examined the mechanism of dust-driven winds based on the failed radiatively accelerated dusty outflow (FRADO) model proposed by Czerny \& Hryniewicz, letting it be responsible for the formation of massive outflow. We calculated the probability of observing the BAL effect from the geometry of outflow which is a function of global parameters of black hole mass ($M_{\bullet}$), Eddington ratio (\eddr), and metallicity (Z). We then compared the results with prevalence of BAL QSOs in a sample of observational data from SDSS. The consistency of our model with the data supports the interpretation of the BAL phenomenon as a result of source orientation, rather than a transitory stage in AGN evolution.

\end{abstract}


\maketitle

\section{Introduction}
BAL QSOs represent a unique and enigmatic class of quasars that exhibit prominent blue-shifted absorption features in their spectra \citep{Lynds_1967, Trump2006} which evidence the presence of strong and massive outflows from the source \citep{Allen2011, Hamann2013}. They exhibit a continuous and wide absorption trough encompassing velocities reaching several times $10^4$ km/s \citep{Risaliti2005, Gibson2009}.
The study of BAL QSOs holds significant importance in understanding the intricate interplay between the accretion disk, black hole, and surrounding environment \citep{Hopkins2009, hamann2019}. 

BAL QSOs are divided into three types based on the absorption lines present in the spectrum \citep{hall2002}. High-ionization BAL quasars (HBAL QSOs), constituting approximately 85\% of BAL quasars, exhibit absorption features only from highly ionised atoms. Low-ionization BAL quasars (LBAL QSOs) have high-ionization absorption plus displaying absorption troughs in low-ionisation ions, making up approximately 15\% of the entire BAL population. FeLBAL QSOs, the rarest subtype among BAL QSOs, exhibit absorption from both high- and low-ionization species, as well as absorption from Iron excited states.

Observationally, a comparison of multiple wavelengths reveals minimal intrinsic distinctions between BALs and non-BALs, except for the tendency of BAL QSOs to exhibit redder ultraviolet continua compared to non-BAL QSOs \citep{Weymann1991, Lewis2003}. BALs are commonly associated with outflows from the accretion disk of active galactic nuclei (AGNs) \citep{murray1995}. The origin of BALs in quasars is subject to two theoretical scenarios. The first suggests that BAL QSOs are essentially normal QSOs observed from a line of sight that passes through the outflowing gas \citep{elvis2000}. The second scenario involves an evolutionary process \citep{williams1999} which is no longer preferred \citep{turnshek88}. 
The first scenario relies on the massive outflow of gas due to line-driving mechanism \citep{elvis2000, risaliti2011}. However, there are shreds of evidence in the spectra of BAL sources hinting at the presence of dust \citep{dunn2010, borguet2013}. We, therefore, aimed at testing whether there is any consistency or correlation between the properties of a dust-driven outflow and the probability of observing BAL feature in a given source. For the comprehensive version of this study, see \cite{NaddafBAL}.
\section{Methodology}

\begin{figure}
  \centering  
  \includegraphics[width=0.69\textwidth]{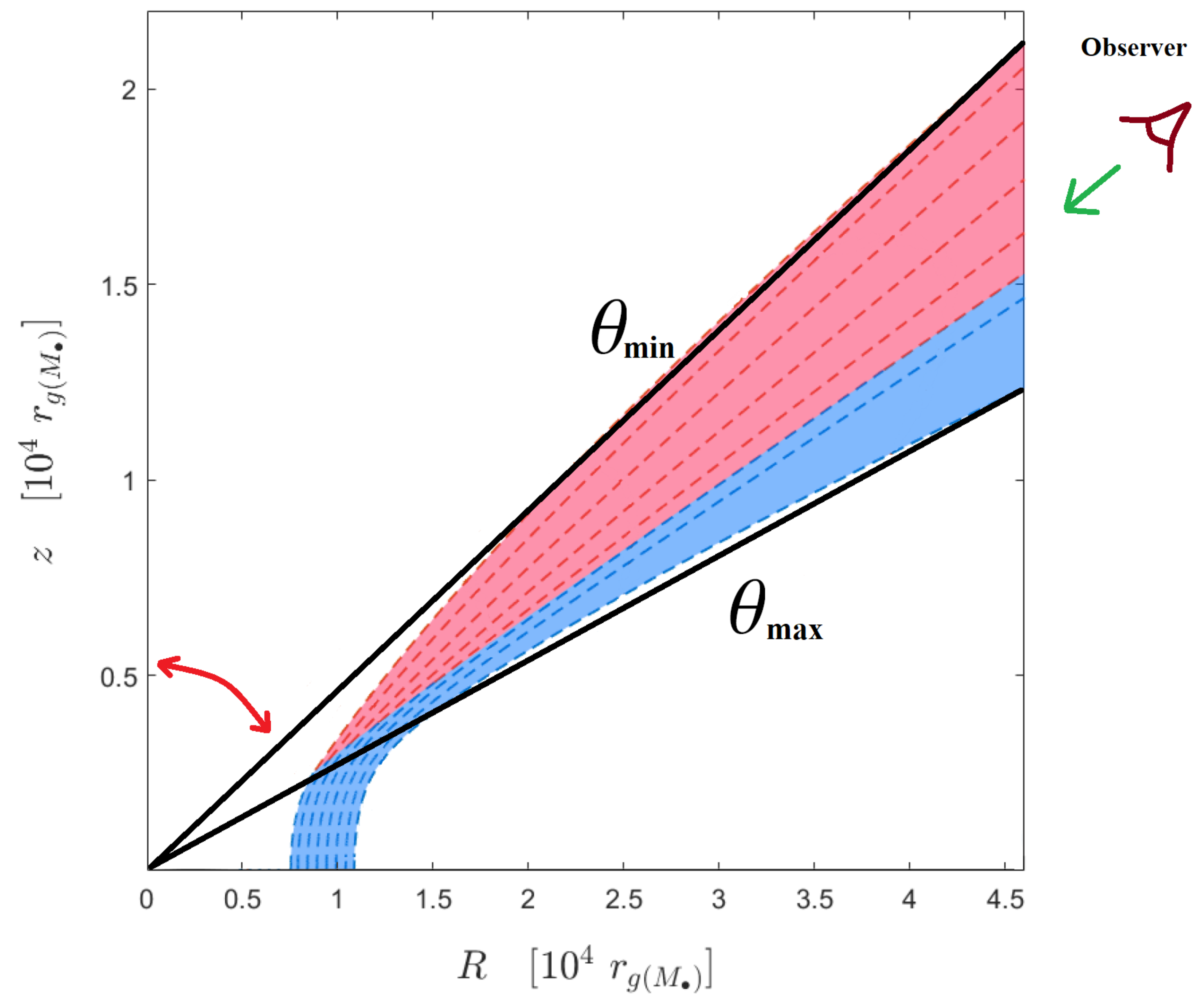}
\caption{An example of outflow in 2.5D FRADO model for a source at Eddington rate with $M_{\bullet}$ of $10^8 M_{\odot}$ and metalicity of $5 Z_{\odot}$. Dotted lines show the path of the trajectory of a clump. The area shaded in blue/red corresponds to the dusty/dustless (lower/higher ionized) actual status of outflowing material. The two black solid lines represent the maximum and minimum angles (to symmetry axis, $z$) confining the outflow cone. The BAL feature is seen if the line of sight is within the outflow cone.}
\label{fig:geometry}
\end{figure}

We use the enhanced 2.5D version of the failed radiatively accelerated dusty outflow model (FRADO) \citep{czerny2011, naddafczerny2022}. FRADO model predicts that radiation pressure acting on dust lifts the clumps of dusty material from the surface of an accretion disk. Depending on the launching location and the global parameters, i.e. black hole mass ($M_{\bullet}$), Eddington ratio (\eddr), and metallicity ($Z$), the material may then come back to the disk surface or it may escape from the gravitational potential. Also, the launched material may or may not lose the dust content depending on the trajectory \citep[for more details of the 2.5D model, see][]{naddaf2021, naddafczerny2022}. An example of the outflow in the 2.5D FRADO model is shown in figure 1 which in this case includes both highly- and lowly-ionized regions shaded in red (dustless) and blue (dusty), respectively. The geometry of the outflow is a function of $M_{\bullet}$, \eddr, and $Z$. We calculate the probability of observing BAL feature as

\begin{equation}
   P_{\rm BAL} = \dfrac
   {{\rm fraction~of~the~sky~covered~by~outflow~cone}}
   {{\rm unit~sphere}}
   =\cos(\theta_{\rm min}) -
     \cos(\theta_{\rm max})
\end{equation}

As for the observational data, we collected a sample from the quasar catalogue of the Sloan Digital Sky Survey DR7 \citep{shen2011}. 
The selected sample consists of sources with a median S$/$N$>10$ per pixel in the rest-frame 2700-2900\AA\ region in which the black hole masses are estimated from \mgii\ . The full sample contains 42,349 objects that 3\% of which are detected as BAL QSOs. Black hole masses and Eddington ratios for \mgiia were obtained from the SDSS catalog.
Dividing the parameter space of black hole mass and Eddington ratio, $M_{\bullet}$-\eddr, into bins centered at integer values of $\log M_{\bullet}$ and $\log$\eddr
with the step size of one in log scale, the median values for the distribution of HBAL and LBAL sources of the sample in each bin are presented in Figure 2. The value from the sample of observational data which can be compared with $P_{\rm BAL}$ from 2.5D FRADO is the {\it prevalence ratio}, $r_{\rm BAL}$, defined as
\begin{equation}
   r_{\rm BAL} =\dfrac
   {N_{\rm BAL}}{(N_{\rm BAL}+N_{\rm nonBAL})}
\end{equation}
where $N$ stands for the corresponding number of sources in each bin.

\begin{figure}
  \centering  
  \includegraphics[width=0.8\textwidth]{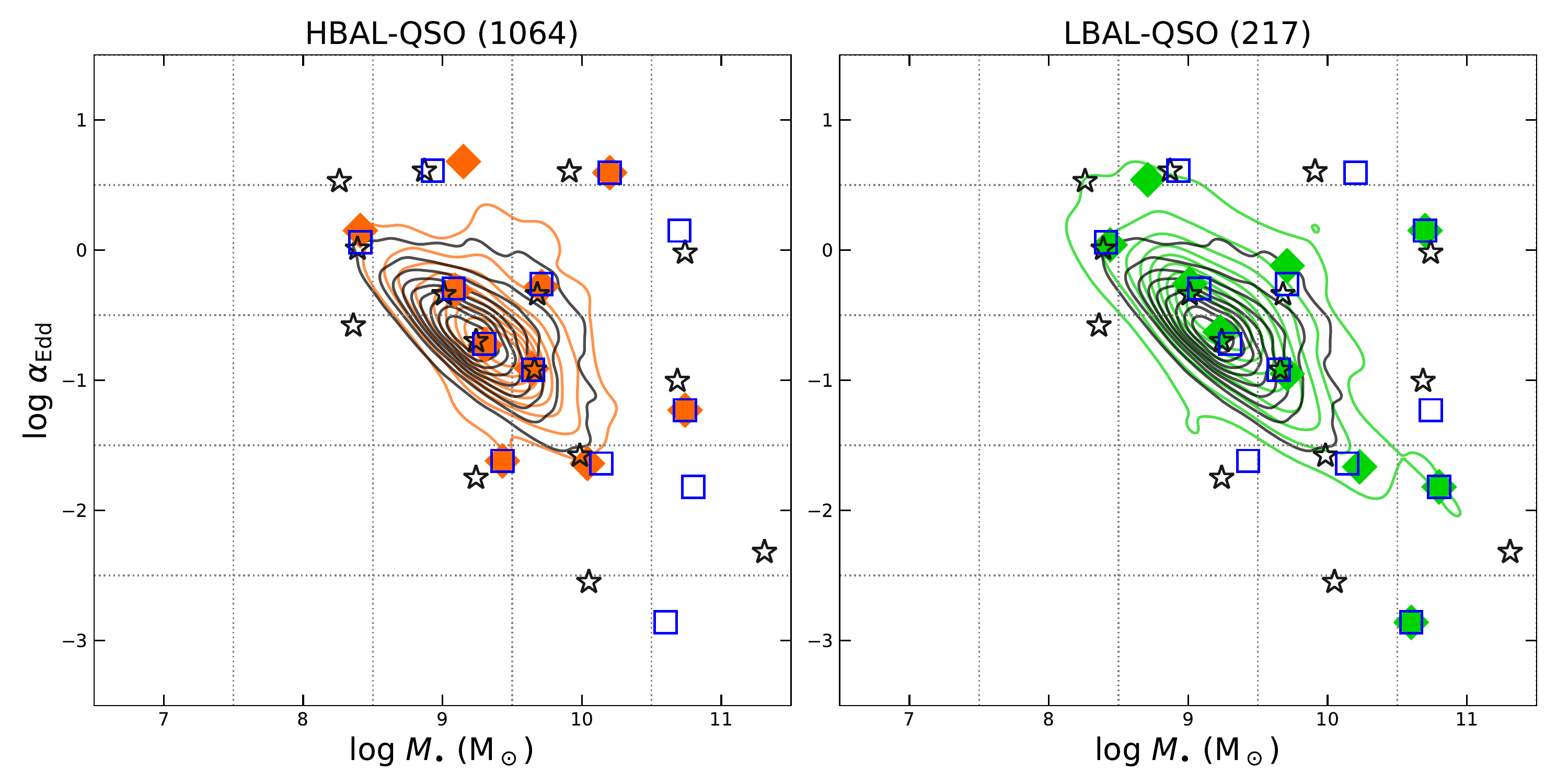}
\caption{The $M_{\bullet}$-\eddr\ space for non-BAL (black $\smallstar$), BAL (blue $\square$), HBAL (orange $\diamond$ in the left panel) and LBAL (green $\diamond$ in the right panel) QSOs. Symbols represent the median values of $\log{M_{\bullet}}$ and log \eddr\ in each bin. Contours in black, orange, and green mark the distribution for non-BAL, HBAL, and LBAL QSOs, respectively.} \label{fig:mbh_edd}
\end{figure}

\begin{figure}
  \centering  
  \includegraphics[width=\textwidth]{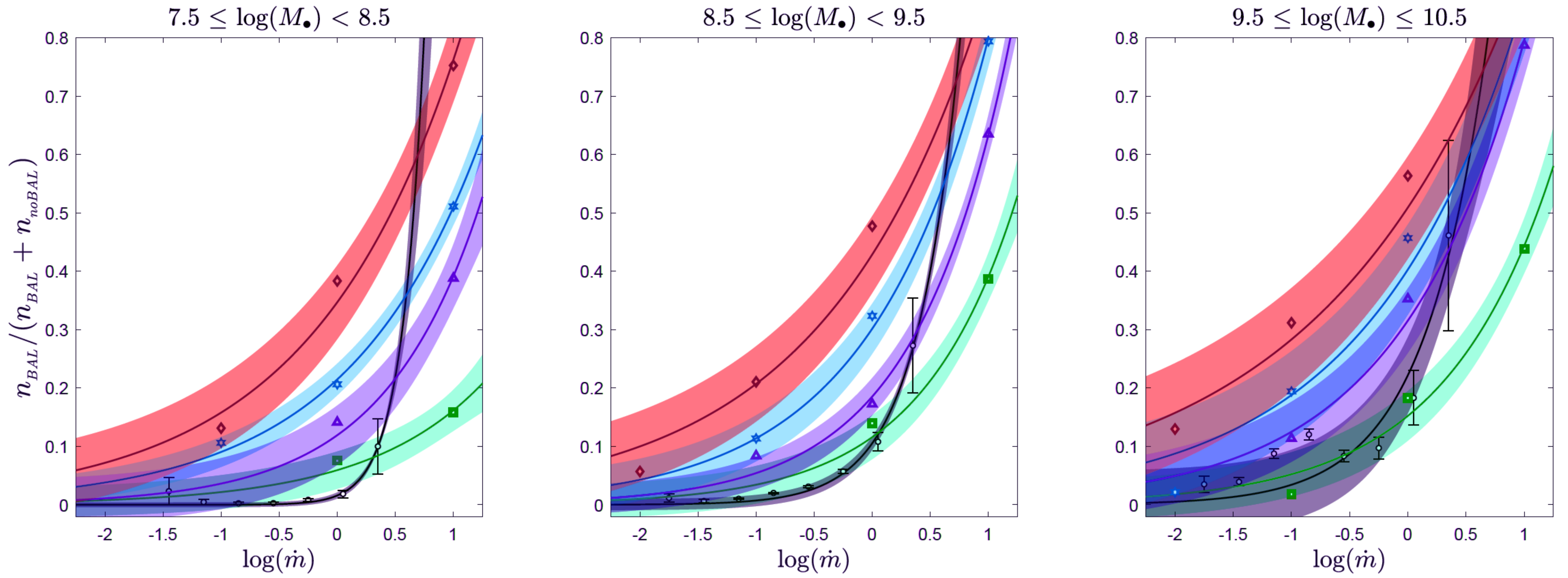}
\caption{Comparison of $r_{\rm BAL}$ from the sample of observational data with $P_{\rm BAL}$ from 2.5D FRADO, for the range of $M_{\bullet}$, and \eddr. Solid lines show the best exponential fit to the corresponding data. The values of $r_{\rm BAL}$ and the corresponding fit in each panel are shown in black. Results from 2.5D FRADO, $P_{\rm BAL}$, for metallicities of 1, 2.5, 5, and 10 $Z_{\odot}$, and their corresponding fits are color coded in green, violet, blue, and red, respectively.}
\label{fig:trends}
\end{figure}

\section{Result \& Discussion}

Utilizing the 2.5D FRADO model, we obtained the geometry of outflow for a range of global parameters of $M_{\bullet}$, \eddr, and $Z$. We then quantified the associated parameter $P_{\rm BAL}$. Concurrently, we analyzed an observational sample to extract the corresponding $r_{\rm BAL}$.
The figure 3 presents a comparative analysis between the simulation results generated by 2.5D FRADO (colored data points) and observational data (black data points).
Remarkably, exponential functions emerged as optimal fits that effectively captured the trends exhibited by the data points in all cases. This striking correlation between theoretical predictions and observational data lends substantial support to the hypothesis that the BAL effect is not a temporary phenomenon but rather arises from the inherent orientation of the source relative to the observer. Specifically, our findings indicate that the BAL effect, both in our model and in the empirical data, is predominantly associated with sources possessing $M_{\bullet}$ exceeding $10^8 M_{\odot}$, and also the effect gets amplified as the accretion rate of the source rises.
Here we presented a subset of our results in which the presence of torus is not taken into account for computing the $P_{\rm BAL}$. In this particular scenario, characterized by the absence of a torus, our analysis reveals a preference for lower metallicities, as evident from the observed trends in the empirical data. However, through further investigations incorporating the torus \citep[see][for details]{NaddafBAL}, we found that higher metallicities are more favored especially when the accretion rate of the source is also higher, as frequently concluded in other studies \citep[see e.g.][]{martinez2018, panda2019b, Panda_etal_2020, Panda2021, sniegowska2021, Panda2022, Garnica_etal_2022}.

\bibliographystyle{iaulike}
\bibliography{iauguide}

\end{document}